\documentclass[prl,twocolumn,showpacs]{revtex4}
\usepackage{hyperref}%,preprintnumbers,amsmath,amssymb
\usepackage[dvips]{graphicx,color}
\begin{document}
\preprint{APS/123-QED}
\title{Emergence of multiple time scales 
in coupled oscillators with plastic frequencies}
\author{Masashi Tachikawa}
\email{mtach@complex.c.u-tokyo.ac.jp}
\author{Koichi Fujimoto}
%\email{fujimoto@complex.c.u-tokyo.ac.jp}
\affiliation{
  Department of Pure and Applied Sciences, College of Arts and Sciences,
   University of Tokyo, 3-8-1, Komaba, Meguro, Tokyo, 153-8902, Japan}
\date{\today}

\begin{abstract}
A coupled-phase oscillator model where each oscillator has an
angular velocity that varies due to the interaction with other oscillators 
is studied. This model is proposed to deepen the understanding of  
the relationship 
between the coexistence and the plasticity of  time scales in complex systems. 
It is found that initial conditions close to a one cluster state 
self-organize into multiple clusters with different angular velocities.
Namely,  hierarchization of the time scales 
emerges through the multi-clustering process in phases.
Analyses for the clusters solution, the stability of the solution and  
the mechanism determining the time scales are reported.
\end{abstract}

\pacs{05.45.+b, 05.45.Xt}

\maketitle

Recently, dynamical phenomena with wide ranges of time scales 
are attracting much attention as these play important roles in many observed 
complex systems in Physics, Chemistry, and  especially  Biology\cite{BD04,KS98}.
For example, reflection, adaptation and  memorization
of reaction behaviors in biological organisms
are identified with different characteristic time scales whose
correlation gives rise to unified responses of individuals.  
Since the time scale diversity is observed not only 
in higher organisms with advanced neural systems but also in single cell
organisms such as bacteria \cite{Kor04} and  paramecia \cite{Nak82}, 
it is thought to be an essential property for all living organisms. 
There is few frameworks 
to discuss the correlation over wide variety of time scales. 
The development of theoretical 
approaches for multiple time scale  systems 
very much remains a topic of current interest \cite{FK03}.

Another remarkable phenomenon related to multiple time scales in 
biological systems is the plasticity of time scales. 
For example, some types of neurons have variable spiking 
frequencies that are thought to be important for characterizing the  
dynamics of neural systems \cite{BD04, Ego02}. 
Additionally, oscillations of metabolic reactions 
have also been reported to display widely changing frequencies 
\cite{Shi95}.  
It can be expected, therefore, that the plasticity of the time scales
has a direct impact on the diversity in these systems.

Since the time scales in biological systems are often 
strongly correlated with concentrations 
that can be modeled as simple variables,
it is  appropriate that the plasticity of time scales is described 
as the change of dynamical variables in models.
For example, an enzyme concentration 
regulates the reaction rate in a Michaelis-Menten type reaction.
Additionally,  the bursting rates of some neuron cells are controlled 
by calcium concentrations  in the cells \cite{Ego02}. 
Therefore, large networks of enzyme reactions or  assemblies of 
neurons can be studied based on the plasticity of the time scales.
For a practical example, 
a model treating frequencies as variables dependent on a 
metabolite accurately reproduces certain behaviors of the 
true slime mold \cite{Tak97} .

\vspace{0.5em}

In this paper, we present a mechanism that can generate
a wide variety of time scales based on time scale plasticity.  
The multiplicity of the time scales is not provided a priori, 
and it is assumed that the plasticity of the characteristic time scales 
depends on the state of the system. 
In the context of  dynamical systems modeling,  this means that
variables need to be introduced which represent 
the characteristic time scales, and that 
the  hierarchy of the time scales is supposed to be generated 
through a self-organizing process \cite{NP77}. 

We focus on models of coupled oscillators which have 
the fruitful behavior in dynamics  
such as entrainment, phase locking, clustering and so on 
\cite{Win67},
and introduce plastic frequencies (i.e. time scales) on them. 
The realization with phase oscillators, 
as one of the simplest case, is given by 
\begin{eqnarray}
\left\{\begin{array}{ll}
\displaystyle \dot{\theta_i} = w_i\\
\displaystyle \dot{w_i} = \frac{1}{N} \sum^N_{j=1} 
\left(1- \cos(\theta_i -  \theta_j) \right) - \alpha w_i \
\end{array}\right. \label{eq:model1}
\end{eqnarray}
where the angular velocity $w_i$ of oscillator $i$   
denotes the inverse of the time scale and is 
a variable that is varied based on the phase differences 
($\theta_i - \theta_j$) in an all-to-all coupling scheme.  
Increasing the phase differences lead to an acceleration of the oscillators,
while global synchronization stops them. The parameter
$\alpha$ is set to $0.1$  in order to assure  that the variance ratio
of $w$ is smaller than that of $\theta$.

This model shows multiple clusters states 
in which clusters with different number of oscillators 
have different frequencies.
Namely, the  hierarchical structures in the frequency space 
emerge through the clustering process of oscillators in phases. 
When such a hierarchy of clusters exists,  
the dynamics of even widely separated time scales 
are essentially correlated 
for the reason that  the stability of each cluster
is maintained by its interaction with other clusters.

\vspace{0.5em}

\begin{figure}
\includegraphics[width=6cm]{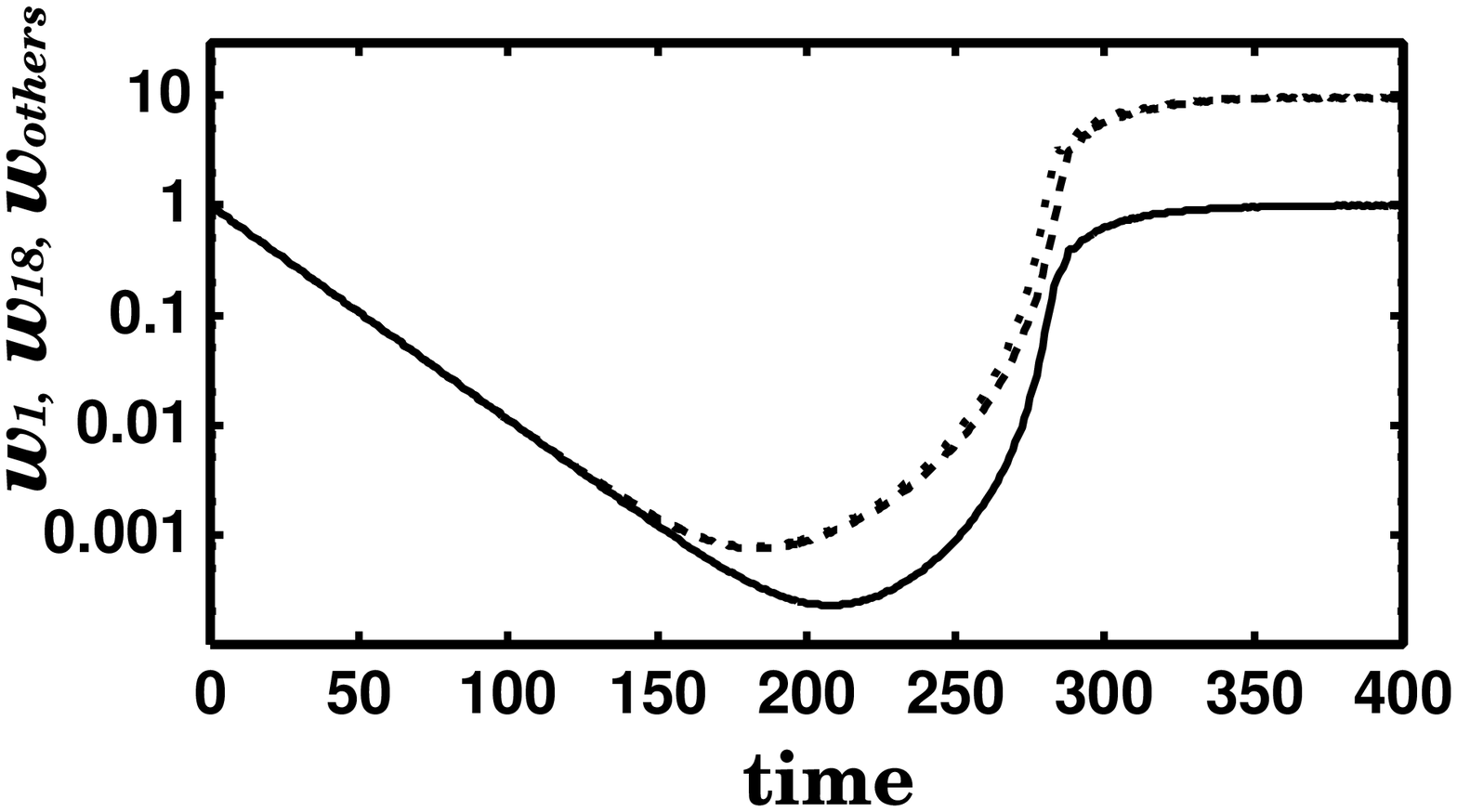}
\includegraphics[width=6cm]{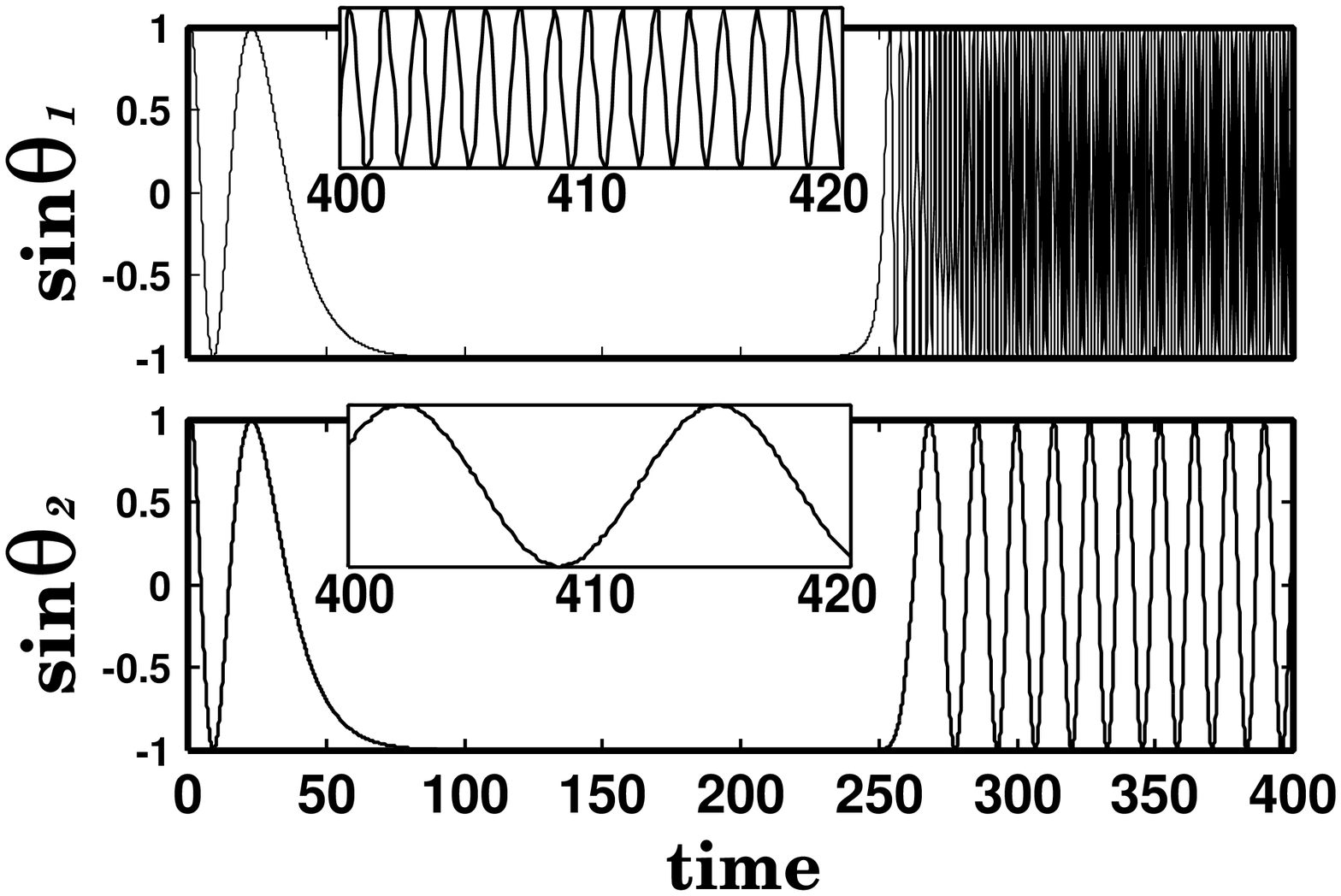}
\caption{
Top: Time series of 
$w_1$ (dotted line), $w_{18}$ (dashed line) and 
$w_{{\rm others}}$ (slid line).
Middle \& Bottom: Time series of 
$\sin(\theta_1)$ and   $\sin(\theta_{2})$.
$N=20,\alpha=0.1$.  
Oscillator 1 and 18 separate to form a smaller cluster, thus 
a 2:18 clustering state emerges. 
The convergence ratio of $w_i$ between the clusters 
agrees with the inverse of the ratio of oscillators numbers 
(\ref{eq:fracfreq}). 
}
\label{fig:1}
\end{figure}

\begin{figure}
\includegraphics[width=5cm]{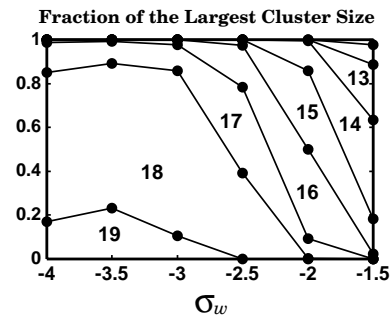}
\caption{
The size of the largest cluster 
is plotted as a function of the variance 
of the initial conditions.
The variance of $\theta$ is fixed to $10^{-2}$ and 
the variance of $w$ is bounded by  $10^{-4}\le\sigma_w\le 10^{-1.5}$ . 
1000 samples are tested for each value. 
The numbers in the figure represent the cluster-sizes obtained and, as can be seen,
initial conditions closer to the one-cluster state yield larger uniform clusters,  broadening
the spread of the time scales. 
}
\label{fig:2}
\end{figure}

Here, we present our numerical results for equation (\ref{eq:model1}). In Fig.\ \ref{fig:1}, it is shown that when taking a
set of $w_i$ and $\theta_i$ near the one-cluster state as initial conditions,  the emergence of a hierarchy in the time scales can be observed 
in a form of symmetry breaking.
Typically the system evolves over time as follows.
First, the frequencies (angular velocities $w_i$) of all oscillators 
approach zero coherently while the phases are synchronized.
Second, a few oscillators separate from the cluster
and start to oscillate rapidly (oscillator $1$ and $18$ in Fig. \ref{fig:1}). 
Then, the oscillators in the remaining cluster attain a uniform frequency and 
phase-synchronization, i.e. a stable cluster is formed.
The fraction of the size of the largest cluster is shown in
Fig.\ \ref{fig:2}.

The oscillators separated from the one cluster state,
in some cases, form smaller clusters to be multiple clusters states,
or oscillate independently in other cases.   
While the one cluster state is unstable and splits spontaneously,
the clusters in the split states are robust against 
small perturbations by noise, and thus 
the multiple clusters states are attractors.

Here we note on a two clusters state which contains $n$ and $m$ 
oscillators $(n>m,~n+m=N)$  as a specific case. 
It is numerically observed to fulfill a relation   
for the averaged frequencies of these clusters 
($\overline{w}_n, \overline{w}_m$) and the number oscillators  
\begin{eqnarray}
\frac{\overline{w}_{n}}{\overline{w}_{m}} =\frac{m}{n}. \label{eq:fracfreq}
\end{eqnarray}
As a result, this very simple model 
generates the robust structures of hierarchical frequencies 
which  fulfill the  simple relation.

The properties of the model can be summarized as follows; 
(i) a single coherent state is unstable and splits into some groups,  
(ii) clusters in  split states are stable,  
(iii) the fraction of frequencies fulfills the condition 
(\ref{eq:fracfreq}) which generates
the hierarchy of time scales from multi-clusters structures.

Since the interacting term only affects the variance of $w_i$, 
items (i) and (ii) concerning the coherence in $\theta_i$
reveal nontrivial properties. These are discussed in detail below. 
First, however, let us consider item (iii) which has a clear mechanism.

The frequencies of the clusters in a steady state are
regulated by the states themselves. 
Since the couplings among synchronized oscillators have no effect, 
the $w_i$ of an oscillator that joins a cluster is activated 
by oscillators that don't join the cluster.
As $\alpha$ is set to $0.1$ and the variation of 
$w_i$ is slower than that of $\theta_i$,
the effect of the coupling 
$1- \cos(\theta_i -  \theta_j)$ 
is averaged over time, and
the averaged frequencies are 
proportional to the number of oscillators
that do not join the clusters.
Therefore, when two clusters with $n$ and $m$ oscillators coexist,
the ratio of the averaged frequencies 
($\overline{w}_{n}/\overline{w}_{m}$) fulfills the condition 
(\ref{eq:fracfreq}). 
Applying this discussion to multi-clusters states, 
where there are $l$ clusters with $n_1,\cdots,n_l$ oscillators 
respectively ($\sum n_i =N$), 
the relation between averaged frequencies is
\begin{eqnarray}
\frac{\overline{w}_{n_i}}{\overline{w}_{n_j}} 
=\frac{N-n_i}{N-n_j}. 
\label{eq:fracfreq m}
\end{eqnarray}
Thus, multi-clusters states leads to multi-frequencies.

\vspace{0.5em}

In order to analyze how the one cluster state becomes unstable 
and how  the multi-clusters states emerge, corresponding to (i)  
and (ii), we now introduce a reduced description of the system.

\begin{figure}
\includegraphics[width=7cm]{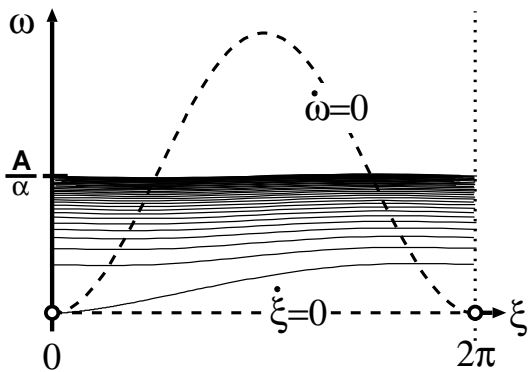}
\caption{
Phase space of the two clusters description in 
relative coordinates. 
Open circles denote the one cluster state, and the dotted line ($\xi = 2\pi$) 
coincides with the line $\xi = 0$ to form the cylindrical phase space.
Null clines (dashed lines) and typical orbits separating from the one cluster state
(thin solid lines) are also plotted . 
The orbit converges around $\omega=A/\alpha$.
}
\label{fig:3}
\end{figure}

To investigate the stability of the one cluster state, 
we take a two-cluster state as the initial condition,
\begin{eqnarray}
\left\{
\begin{array}{l}
w_i=w_{f},~~\theta_i=\theta_{f}~~~~i=1,\dots ,m  \\
w_j=w_{s},~~\theta_j=\theta_{s}~~~~j=m+1,\dots ,N
\end{array}\right\} \\
(N=m+n,~~~n > m )~~~~~~~~ \nonumber
\end{eqnarray}
so that the system reduces to four variables.
If we then introduce the relative coordinate 
\begin{eqnarray}
\omega = w_i-w_j,~~~~\xi = \theta_i-\theta_j ,
\end{eqnarray}
the description can be further reduced to only two variables 
in cylindrical phase space,
\begin{eqnarray}
\left\{\begin{array}{ll}
\dot{\xi}    = \omega \\
\dot{\omega} = A\{1- \cos(\xi)  \} - \alpha \omega 
\end{array}\right.\label{eq:2cls_rel} \\
A \equiv \frac{n-m}{N}.~~~~~~~~~~~~~~~~~~
\end{eqnarray}
The phase space of the $(\xi, \omega)$ system and the null cline
are shown in Fig.\ \ref{fig:3}, where 
the origin corresponds to the one cluster state
\footnote{
The following discussion does not depend on any specific values of
 $n,~m$ and hence its results are valid for any ratio.}.

Expanding equation (\ref{eq:2cls_rel}) at the origin, we obtain
\begin{eqnarray}
\left\{
\begin{array}{l}
\dot{\xi} = \omega  \\
\dot{\omega} =  - \alpha\omega + A\xi^2 +O(\xi^4)
\end{array} \right.  .
\end{eqnarray}
The vector field at the origin 
has a zero and a negative eigenvalue,
but when considering the second order we find that 
the origin, 
the one cluster state, is unstable.

The stable solution of the two clusters state 
$\overline{\omega} =\omega_{\alpha}(\xi)$, 
which is a limit cycle circling around the cylinder,  
is calculated as follows:
When averaging $\cos(\xi)$ in (\ref{eq:2cls_rel}),
a limit cycle exists around the averaged line 
$\overline{\omega} = \frac{A}{\alpha}$, 
and $d \omega/d \xi \sim \alpha \ll \overline{\omega}$
indicates that the variation from the average
is small (see Fig.\ \ref{fig:3}).
Then, by substituting the solution expanded in $\alpha$ 
into $d \omega/d \xi = \dot{\omega}/\dot{\xi}$ 
and by comparing the both sides, we obtain 
\begin{eqnarray}
\omega = \frac{A}{\alpha} - \alpha \sin(\xi) 
+ O(\alpha^3). \label{eq:sol 2dim}
\end{eqnarray}
which is in fairly good accordance with the
 numerical simulation.

\vspace{0.5em}

In order to discuss the stability of the clusters in the two clusters state,
we divide a cluster of them  virtually and again describe the 
displacements with relative coordinates. 
Suppose a  cluster containing $n$ oscillators is divided into 
two clusters with $n_1,~n_2$ oscillators ($n=n_1+n_2$).
Let $\epsilon$ and $\delta$ be the differences of the phases and 
the frequencies between the clusters respectively.
Because our concern is the linear stability of the cluster, 
we focus on the region with sufficiently small $\epsilon$ and $\delta$. 
The previously employed coordinates $(\xi,\omega)$ are then naturally 
defined as the differences between 
the center of the divided clusters and the other cluster,
\begin{eqnarray}
& & \left\{
\begin{array}{l}
\dot{\xi} = \omega  \\
\dot{\omega} = A\left\{1-\cos(\xi) \cos(\frac{\delta}{2}) \right\}\\ 
~~~~~+ B\sin(\xi)\sin(\frac{\delta}{2})
-\frac{n}{2N}\{1-\cos(\delta) \}-\alpha \omega  \\
\end{array} \right. \label{eq:3clusters 1}\\
& & \left\{
\begin{array}{l}
\dot{\delta} = \epsilon  \\
\dot{\epsilon} = - \frac{2m}{N}\sin(\xi)
\sin(\frac{\delta}{2})   
 + B \{ 1 - \cos(\delta)\}  - \alpha \epsilon 
\end{array} \right. \label{eq:3clusters 2} \\
& & B\equiv \frac{n_2-n_1}{N}
\end{eqnarray} 
The origin of the $(\delta,\epsilon)$ plane corresponds to 
 the two clusters state. 
The region $|\delta|,|\epsilon| \ll 1$ is considered, 
and the behaviors of  $\xi$ and  $\omega$ are 
 approximated by the solution of (\ref{eq:sol 2dim}). 
Linearizing (\ref{eq:3clusters 2}) at the origin, we obtain
\begin{eqnarray}
\left\{
\begin{array}{l}
\dot{\delta} = \epsilon  \\
\dot{\epsilon} = - \frac{m}{N}\sin(\xi)\delta - \alpha \epsilon 
\end{array} \right. \label{eq:slow 2} .
\end{eqnarray} 
Here, $\sin(\xi)$ can be treated as a forced oscillation term 
that is approximated as
\begin{eqnarray}
\xi &\simeq& \frac{A}{\alpha}t
+\frac{\alpha^2}{A} \cos\left(\frac{A}{\alpha} t\right)  \nonumber \\
\sin(\xi) 
&\simeq& \frac{\alpha^2}{2A} +\sin \left(\frac{A}{\alpha} t\right) + 
\frac{\alpha^2}{2A} \cos\left(\frac{2A}{\alpha} t  \right),  
\label{eq:oscillate} 
\end{eqnarray}
where $\dot{\xi}=\omega\simeq\frac{A}{\alpha}-\alpha \sin(\xi)$ 
is integrated by using successive approximations.

If a normal averaging method is applied, 
only the first constant term in (\ref{eq:oscillate}) is taken
to estimate the stability of (\ref{eq:slow 2}). 
However, the amplitude of the oscillation in second term
is much larger than that of the first term (in the order of the square of 
$\alpha$), and hence both terms should be included.
Thus
equations (\ref{eq:slow 2}) can be expressed as  
\begin{eqnarray}
\ddot{\delta} + \alpha \dot{\delta} 
+ \frac{m}{N} \left(\frac{\alpha^2}{2A}+ \sin(\frac{A}{\alpha} t)\right)
\delta=0.
\end{eqnarray}
Setting $\delta=R(t)\exp\{-\frac{\alpha}{2} t\}$, we obtain
the Mathieu equation, 
\begin{eqnarray}
\ddot{R} + (p  -2 q \sin(2 t))R =0\\
\left\{
\begin{array}{l}
p =\frac{\alpha^4}{A^2}\left(\frac{2m}{AN} -1 \right)
\\
q = \frac{2\alpha^2m}{A^2N}
\end{array}\right. 
\end{eqnarray}
Here, $p=O(\alpha^4)\ll 1,~ q=O(\alpha^2)\ll 1$ 
is in the area of Mathieu's parameters  
where $R=0$ is stable.                     
Since $\delta$ is the product of $R$ and the
exponential decay term $\exp\{-\frac{\alpha}{2} t\}$,
$\delta$ is  not destabilized by the $O(\alpha^2)$ 
perturbation which is omitted to approximate $\sin(\xi)$.
Therefore the clusters in the two clusters state can be concluded 
to be stable.

\vspace{0.5em}

In the present paper
we introduce a novel type of coupled oscillator model 
where the time scales occur as variables. It is shown 
that multiple clusters with 
different time scales emerge through a self-organization process based on time scale plasticity. 
This process is characterized by symmetry breaking due to
separation from the one-cluster state.

Though we introduce our model (\ref{eq:model1}) as one of the simplest 
model to discuss time scale plasticity, 
which is also derived from a limit-cycle model 
with some assumptions, 
Suppose we have globally coupled limit cycles of  normal form type 
in polar coordinates 
\begin{eqnarray}
\left\{\begin{array}{l}
\dot{\theta_i} = b(1 - r_i^2) +\frac{1}{N}\sum 
\frac{r_j}{r_i}\sin(\theta_j-\theta_i)\\
\dot{r_i}= a r_i(1-r_i^2) +\frac{1}{N}\sum 
\left(\cos(\theta_j-\theta_i)r_j -r_i\right) ,
\end{array}\right. 
\end{eqnarray}
where the coordinate is chosen that 
the angular velocity becomes zero at the limit cycle. 
Then we assume $b\gg a \gg D$, 
transform $r_i=1-O(D/a)\equiv 1- w_i$ and 
omit the terms of  $O(D/b)$ and $O(D^2/a^2)$,
\begin{eqnarray}
\left\{\begin{array}{l}
\dot{\theta_i} = - 2 b w_i \\
\dot{w_i}= -2 a w_i + \frac{D}{N}\sum \left(
1 - \cos(\theta_j-\theta_i)  \right)
\end{array}\right.
\end{eqnarray}
is obtained. 
Lastly  with appropriate transformations,  
equation (\ref{eq:model1}) is derived.
In this setup, the time scale of the each oscillator diverges at 
the limit cycle, and the plasticity results from the steep 
gradient of frequencies ($b\gg D$) around the cycle.
In this way, such singular properties can also 
lead the plastic time scales \cite{Tac03}.

\begin{figure}
\includegraphics[width=7cm]{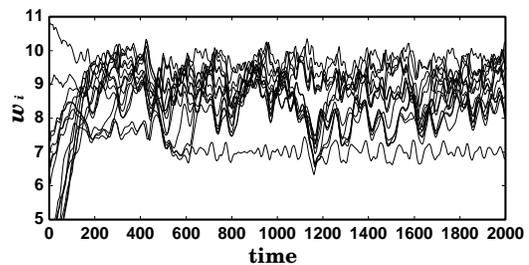}
\caption{
Time series data of $\omega_i$ from a highly spread initial condition.
Neither a large cluster nor large differences between $w_i$
 emerge and the whole system continues to be chaotic.
}\label{fig:4}
\end{figure}

Another type of multiple frequency behavior in coupled oscillators is studied 
in the context of  antenna and radar systems in \cite{In03} where
the hierarchy is characterized by a discrete symmetry group of the system. 
Although those hierarchies  and the ones reported here have 
different mechanisms, 
a unified explanation should be developed for these structures.
Since we use the simplest oscillator as the dynamical element, 
a hierarchy with only a few layers is observed. 
In more complex systems of chaotic oscillators,  
multiple layers can be self-organized 
(to be reported in a forthcoming paper \cite{FT04}).
In this study, the  initial conditions are restricted to the vicinity of the 
one-cluster state. When we choose more scattered out  initial conditions, 
coherent structures do not emerge, chaotic behaviors persist 
and large time scale differences are not generated 
(Fig.\ \ref{fig:4}). 
This implies the importance initial condition selection, 
a requirement  common to many self-organization phenomena 
in highly  complex systems \cite{FK00}.

\vspace{0.5em}

The authors are grateful to A. Awazu, H. Daido, T. Ikegami, K. Kaneko, 
S. Sasa, F. H. Willeboordse and H. Yamada for useful discussions.
This work is supported by a Grant-in-Aid for Scientific Research from 
the Ministry of Education, Science, and Culture of Japan.


\begin{thebibliography}{99999}

\bibitem{BD04}  G. Buzsaki and A. Draguhn, Science, {\bf 304}, 1926 (2004). 

\bibitem{KS98} J. Keener and J Sneyd, {\it Mathematical Physiology},
(Springer-Verlag, 1998);
D. Fell, {\it Understanding the Control of Metabolism},
(Portrand Press, 1997).

\bibitem{Kor04} E. Korobkova et al., Nature, {\bf 428}, 574, (2004).

\bibitem{Nak82} Y. Nakaoka et al., Proc. Jpn. Acad., {\bf 58}, 213 (1982).

\bibitem{FK03} K. Fujimoto and K. Kaneko, Physica D {\bf 180}, 1 (2003);
      K. Fujimoto and K. Kaneko, Chaos {\bf 13}, 1041 (2004).

\bibitem{Ego02} A. V. Egorov et al., Nature, {\bf 420}, 173 (2002).

\bibitem{Shi95} 
T Shinjo et al., Physica D, {\bf 84}, 212 (1995); 
Y. Adachi et al., J. Immunol, {\bf 163}, 4367 (1999);
A. J. Rosenspire et al., Biophys. J., {\bf 79}, 3001, (2000);
L. B. Dale et al., J. Biol. Chem. {\bf 276}, 35900, (2001);
L. F. Olsen et al., Biophys. J., {\bf 84}, 69, (2003).

\bibitem{Tak97} A. Takamatsu et al., 
        J. Phys. Soc. Jpn. {\bf 66} 1638 (1997). 



\bibitem{NP77}   G. Nicolis and I.  Prigogine, 
        {\it Self- organization in nonequilibrium systems},  
          (New York: John Wiley \& Sons,  1977). 

\bibitem{Win67}
A. T. Winfree, {\it The Geometry of Biological Time},
              (Springer, New York, 1980); 
Y. Kuramoto, {\it Chemical Oscillations, Waves, and Turbulence},
              (Springer, Berlin, 1984);
K. Kaneko and I. Tsuda, {\it Complex Systems :  Chaos and Beyond}
(Springer, Tokyo, 2001).



\bibitem{Tac03} M. Tachikawa, Prog. Theor. Phys., {\bf  109}, 133 (2003). 


\bibitem{In03} V. In et al., PRL, {\bf 91}, 244101 (2003). 


\bibitem{FT04} K. Fujimoto and  M. Tachikawa, {\it in preparation}.

\bibitem{FK00} C. Furusawa and  K. Kaneko, 
J. Theor. Biol., {\bf 209}, 395, (2001).




\end{thebibliography}
\end{document}